\documentclass[aps,pra,reprint,showpacs,amssymb,amsmath,superscriptaddress]{revtex4-1}

\usepackage{graphicx}
\usepackage{times}
\usepackage{color}
\usepackage{siunitx}

\providecommand{\abs}[1]{\lvert#1\rvert}

\newcommand{\rTF}{\rho_\mathrm{0}}

\newcommand{\ket}[1]{\lvert#1\rangle}

\newcommand{\Erec}{E_\mathrm{rec}}
\newcommand{\dPsi}{\delta\hat{\psi}}
\newcommand{\Bessel}{\varphi}
\newcommand{\0}{$\ket{0}$}
\newcommand{\rme}{\mathrm{e}}
\newcommand{\Hex}{H_{\mathrm{ex}}}
\newcommand{\Einstab}{E^\pm_{nl}} 
\newcommand{\figref}[1]{Fig.~\ref{#1}}
\newcommand{\llat}{\lambda_\mathrm{lat}}
\newcommand{\expec}[1]{\langle #1 \rangle}
\newcommand{\Rb}{$^{87}$Rb }
\newcommand{\braopket}[3]{\langle#1|#2|#3\rangle}

\begin{document}
\title{Spin dynamics in a two dimensional quantum gas}
\author{Poul L. Pedersen}
\affiliation{QUANTOP, Institut for Fysik og Astronomi, Aarhus Universitet, Ny Munkegade 120, 8000 Aarhus C, Denmark.}
\author{Miroslav Gajdacz}
\affiliation{QUANTOP, Institut for Fysik og Astronomi, Aarhus Universitet, Ny Munkegade 120, 8000 Aarhus C, Denmark.}
\author{Frank Deuretzbacher}
\affiliation{Institut f\"ur Theoretische Physik, Leibniz Universit\"at Hannover, Appelstra\ss e 2, 30167 Hannover, Germany.}
\author{Luis Santos}
\affiliation{Institut f\"ur Theoretische Physik, Leibniz Universit\"at Hannover, Appelstra\ss e 2, 30167 Hannover, Germany.}
\author{Carsten Klempt}
\affiliation{Institut f\"ur Quantenoptik, Leibniz Universit\"at Hannover, Welfengarten 1, 30167 Hannover, Germany.}
\author{Jacob F. Sherson}
\affiliation{QUANTOP,  Institut for Fysik og Astronomi, Aarhus Universitet, Ny Munkegade 120, 8000 Aarhus C, Denmark.}
\author{Andrew J. Hilliard}
\affiliation{QUANTOP, Institut for Fysik og Astronomi, Aarhus Universitet, Ny Munkegade 120, 8000 Aarhus C, Denmark.}
\author{Jan J. Arlt}
\affiliation{QUANTOP, Institut for Fysik og Astronomi, Aarhus Universitet, Ny Munkegade 120, 8000 Aarhus C, Denmark.}

\date{\today}

\begin{abstract}
 We have investigated spin dynamics in a 2D quantum gas. Through spin-changing collisions, two clouds with opposite spin orientations are spontaneously created in a Bose-Einstein condensate. After ballistic expansion, both clouds acquire ring-shaped density distributions with superimposed  angular density modulations. The  density distributions depend on the applied magnetic field and are well explained by a simple Bogoliubov model. We  show that the two clouds are anti-correlated in momentum space. The observed momentum correlations pave the way towards the creation of an atom source with non-local Einstein-Podolsky-Rosen entanglement.
\end{abstract}
\maketitle
Since the optical trapping of Bose-Einstein condensates (BECs) enabled the investigation of quantum gases with multiple spin components,  spinor condensates have become a particularly rich research field~\cite{stamper-kurn13,Kawaguchi2012253}. While initial work focused on an understanding of the ground state and dynamical properties of spinor condensates~\cite{stenger98,chang04,widera05}, recent experiments have started to exploit their properties for applications in other fields. In particular, the production of entangled states through spin dynamics~\cite{gross11,hamley12} has spawned interest in spin dynamics for their applications in precision metrology~\cite{luecke11,Gross2010}.

Spin dynamics in a trapped  quantum gas are strongly influenced by  the geometry of the confining potential. In particular, highly asymmetric optical traps  provide a  way to reduce the dimensionality of a trapped quantum gas, both with respect to the motional and spin degrees of freedom~\cite{sadler06}. Thus, tailored confining potentials offer new avenues for exploiting spin dynamics, e.g.,  the generation of correlated pairs of atoms in well-defined motional states \cite{scherer13}, similar to work on four-wave-mixing of ultra-cold atoms in an optical lattice \cite{hilligsoe05,campbell06,bonneau13}. 

In this Rapid Communication, we investigate spin dynamics in a quantum gas confined to two dimensions (2D) by an optical lattice. We show how the spin excitation modes in the 2D  potential lead to ring-shaped density distributions with a superimposed angular density modulation in time-of-flight images. The angular structure is traced  to the matter-wave interference  between multiple spin excitation modes with angular momentum.  The observed density distributions may also be interpreted as several wave packets propagating in 2D with well-defined  momentum. 

We investigate spin dynamics in a \Rb BEC prepared in  \mbox{$\ket{F=2,m_F=0}$} ($\ket{0}$). By making several standard approximations to treat  atomic collisions at ultra-low temperatures,   one finds that only collisions that preserve the total magnetization can occur~\cite{stamper-kurn13}. 
Thus, the spin dynamics lead to scattering into \mbox{$\ket{F=2,m_F=\pm 1}$} ($\ket{\pm1}$) and \mbox{$\ket{F=2,m_F=\pm 2}$} ($\ket{\pm2}$), but for short evolution times, scattering between \0 and $\ket{\pm1}$ predominates; i.e., \mbox{$\ket{0}+\ket{0} \leftrightarrow \ket{1}+\ket{-1}$}. By treating the $\ket{0}$ condensate as a classical field $\psi_0$ and $\ket{\pm1}$ as small fluctuations $\dPsi_{\pm1}$,
 the dynamics may be  described by
  \begin{align}
    &\hat{H} = \int d^2\mathbf{r}\,\psi_0^* \left(\hat{H}_0+ \frac{U_0}{2} n_0 - {\mu}\right) \psi_0\nonumber\\
    &+ \sum_{m=\pm1} \int d^2\mathbf{r}\, \dPsi_{m}^\dagger \left(\hat{H}_0 + (U_0 + U_1)n_0 - {\mu} + q\right)\dPsi_{m}\nonumber\\
    &+ U_1 \int d^2\mathbf{r}\, n_0 \left( \dPsi_1^\dagger \dPsi_{-1}^\dagger + \dPsi_1 \dPsi_{-1} \right),
    \label{eq:hamiltonian1}
  \end{align}
where $\hat{H}_0=-\frac{\hbar^2}{2M}\nabla^2 + \frac{M\omega_\rho^2}{2}\rho^2 $, $n_0=\abs{\psi_0}^2$, $\rho$ is the radial coordinate, $\omega_\rho$ is the radial trapping frequency of the confining potential, and $M$ is the mass. The 2D interaction energies  $U_0 $ and $U_1$ respectively describe spin independent and dependent contact collisions~\cite{deuretzbacher10}. The 2D chemical potential is given by ${\mu}$ and  $q\simeq-72~{\mathrm{Hz}}{\mathrm{G}^{-2}}\times B^2$ is the quadratic Zeeman energy  difference between $\ket{0}$ and $\ket{\pm1}$ for a magnetic field of magnitude $B$. The first two lines of Eq.~\eqref{eq:hamiltonian1} describe the internal dynamics of  $\psi_0$ and  $\dPsi_{\pm}$, respectively, while the final line describes scattering between  $\ket{0}$ and $\ket{\pm1}$. 
 Given our focus on spin-changing collisions during short evolution times, we neglect  the significantly weaker  dipolar interaction: in $F=2$ of $^{87}$Rb, the magnetic dipole interaction strength is on the order of 5\% of the spin-dependent contact interaction $U_1$ \cite{stamper-kurn13,vengalattore08}.
\begin{figure}[htb]
  \includegraphics[width=8.6cm]{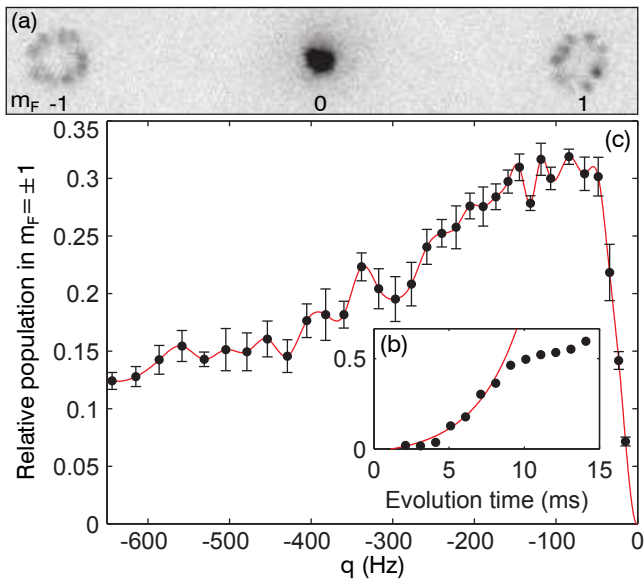}
	\caption{(Color online) (a) Absorption image after Stern-Gerlach separation and time-of-flight for $q=-330~\mathrm{Hz}$ and 8~ms evolution time. (b) Time evolution of relative population in $\ket{\pm1}$. The data have been fitted with an exponential function for the first 8~ms; the statistical uncertainty of each point is on the order of the symbol size. (c) Mean relative population in $\ket{\pm1}$ after 8~ms evolution time;	the solid line is a guide to the eye.}
	\label{fig:spectrum}
\end{figure}

We  realize a 2D spinor gas by preparing a BEC in an  optical lattice. The BEC is produced in $\ket{+2}$  in  
 a Ioffe-Pritchard magnetic trap and contains on average $\sim2\times10^5$ atoms with no discernible thermal fraction ~\cite{bertelsen07}. The BEC is loaded into a red-detuned vertical optical lattice at wavelength \mbox{$\llat=914$~nm} and  depth \mbox{$s \equiv V_\mathrm{lat}/\Erec =18.2\pm0.3$}, where $\Erec$ is the recoil energy. The lattice is formed by retro-reflecting a beam with a $1/\rme^2$ waist of  $w=102\pm2~\mu$m. At  the chosen lattice depth, the radial and axial  trapping frequencies are $\omega_\rho=2\pi\times(47.3\pm1.0)$~Hz and  $\omega_\mathrm{ax}=2\pi\times(23.5\pm0.2)$~kHz respectively, and tunneling between lattice sites is negligible. 
The lattice loading is performed by simultaneously reducing the current in the magnetic trap coils and increasing the lattice intensity over 110~ms. This results in $\sim10$ lattice sites occupied by independent BECs.
The atoms are prepared in \0 by two sequential microwave pulses via \mbox{$\ket{F=1,m_F=1}$} at a constant magnetic bias field of 285~mG in a horizontal direction, which is maintained throughout the experiment.
Following preparation in the lattice, the cloud has a finite thermal component: on average, we obtain $N_\mathrm{BEC}=(4.2\pm0.7)\times10^4$ atoms in the BEC and  $N_\mathrm{th}=(1.2\pm0.2)\times10^5$ in a thermal fraction at temperature $\sim 120$~nK.  The thermal component arises from a technical limitation in  switching off our magnetic trap.
Spin dynamics are initiated by applying an additional magnetic field along the vertical direction. 
This field is turned on with a linear ramp over 1~ms  and adds vectorially to the  magnetic bias field 
to set  $q$; the resulting magnetic field is held constant for a variable evolution time. 
The spin dynamics are brought to an end by switching off the optical lattice. 
 
We probe the result of the spin dynamics by Stern-Gerlach separation and absorption imaging along the vertical direction after 20~ms  time-of-flight.
To avoid saturating the optical depth of our imaging system
 for the $\ket{0}$ cloud, a third microwave pulse is applied just after the lattice is switched off to transfer part of the \0 population to $\ket{F=1, m_F = 0}$, which is transparent to the imaging light~\cite{Ramanathan2012}. 
To obtain accurate atom numbers, the imaging system was calibrated following  \cite{reinaudi07,gajdacz:083105}.
Finally,  the point-spread function of the imaging system is well-described by a Gaussian function with  $1/\rme^2$ waist  5.72~$\mu$m.
A typical absorption image is shown in \figref{fig:spectrum}(a). For this image the peak optical depth of atoms in state $\ket{0}$ was reduced to 1.5. The $\ket{\pm1}$ clouds have an interesting ring structure with a number of peaks on the circumference and will be discussed in detail below.

Figure~\ref{fig:spectrum}(b) shows the time evolution of the population in $\ket{\pm1}$ for $q = -67$ Hz 
in terms of the relative population $(N_{-1}+N_{1})/(N_{-1} + N_0 + N_1)$. The data have been fitted  by an exponential function $\propto \exp(t/\tau)$ from 0-8~ms. This dependence is expected because Eq.~\eqref{eq:hamiltonian1} describes an initial exponential amplification analogous to parametric down conversion in non-linear optics \cite{klempt10}. The fitted value of the time constant was  $\tau=3.0\pm0.2$~ms. After 8~ms, the population deviates from exponential growth, indicating the breakdown of the  `linear regime' in which depletion of $\ket{0}$  can be neglected. Additionally, we begin to observe a small population in $\ket{\pm2}$ at this time: 
at 8~ms, the relative population in $\ket{\pm2}$ is on average 0.25\%, i.e., on the order of our resolution limit. 

\begin{figure}[t]
  \includegraphics[width=8.6cm]{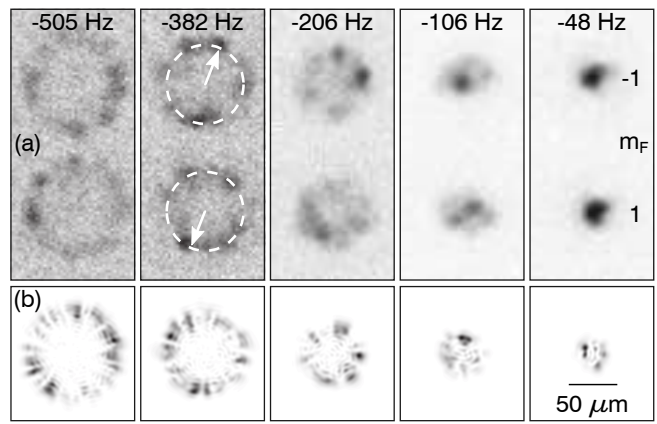}
	\caption{(a) Absorption images for several values of $q$; the  gray-scale  was adjusted  
	for different settings of $q$	to enhance visibility. The schematic for  $q=-382$~Hz illustrates momentum conservation in a spin-changing collision. (b) Simulated time-of-flight  density distributions for the same values of $q$.}
	\label{fig:images}
\end{figure}
The spin dynamics that transfer atoms from \0 to $\ket{\pm1}$ exhibit a clear dependence on $q$, as shown in Fig.~\ref{fig:spectrum}(c). 
 The data show a broad peak centered at $q=-100$~Hz and for higher $\abs{q}$, the population in $\ket{\pm1}$ drops gradually to zero. The $q$-dependence is also evident in the density  distributions after time-of-flight. Figure \ref{fig:images}(a) shows absorption images along the vertical axis for several values of $q$. 
The spatial structure of the clouds  changes from a singly-peaked distribution at low  $\abs{q}$ to a ring-shape  with density modulations around the circumference at higher  $\abs{q}$.

Typically, ring-shaped density distributions in such time-of-flight images indicate the presence of orbital angular momentum~\cite{Ryu2007}, and the appearance of density modulations around the circumference could be interpreted as the matter-wave interference pattern of two or more angular momentum eigenstates~\cite{scherer13,PhysRevA.86.013629}.
For large $\abs{q}$, one may obtain a good understanding of the ring formed in a simplified,  `free-space', picture. In this picture,  one can regard the spin-changing collision as scattering two atoms from the stationary BEC into $|\pm 1\rangle$ in a ring in momentum space with radius $p_\mathrm{rms}=\sqrt{2Mq}$. Momentum conservation requires that the two scattered atoms propagate in opposite directions, as indicated  in \figref{fig:images}(a) for $q =  -382$~Hz. Although, in principle, the system exhibits polar symmetry, bosonic stimulation breaks the symmetry leading to azimuthal modulations, which result in the formation of counter-propagating wave packets. Interestingly, these wave packets should comprise an Einstein-Podolsky-Rosen (EPR) entangled pair in momentum $|\pm p_\mathrm{rms}\rangle$ and spin state $|\pm 1\rangle$~\cite{PhysRev.47.777,PhysRev.108.1070}.

\begin{figure}[tbp]
  \includegraphics[width=8.6cm]{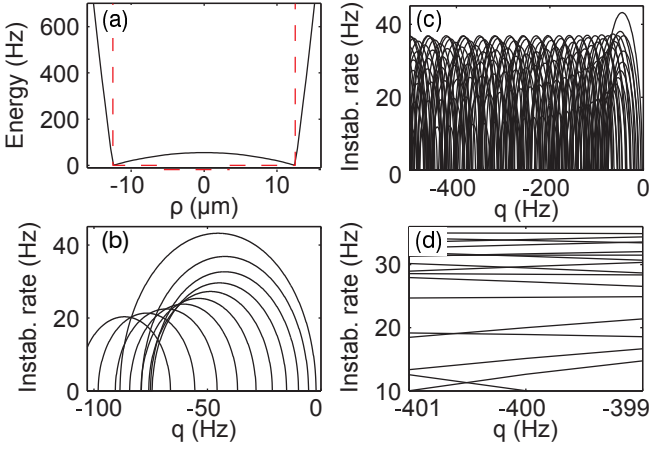}
	\caption{(Color online) (a) (Solid line) Effective potential experienced by $\ket{\pm1}$ and (red dashed  line) the cylindrical box approximation. (b) Instability rates $\Einstab/h$ for $n=1,l=0,1,\dots,9$. The peak instability rate is that of the $n=1,l=0$ mode; the peak  rate decreases monotonically with increasing $l$. (c) Instability rates in the range $q=0$ to $-500$~Hz.
	(d) A close-up of (c) showing the instability rates at $q=-400$~Hz. 
	}
	\label{fig:theory}
\end{figure}

To elaborate on the qualitative picture presented so far, we investigate the excitation spectrum of the $\ket{\pm1}$ states that arises from the
Hamiltonian in Eq.~\eqref{eq:hamiltonian1}. 
In the Thomas-Fermi approximation, $n_0$ in each lattice site has the  shape of the inverted confining potential. Accordingly, atoms in $\ket{\pm1}$ experience  a flat potential bottom plus a small parabolic repulsion $U_1 n_0$~\cite{scherer10}.  Outside the \0 BEC, the potential rises steeply and the effective potential may be approximated by a cylindrical box~\cite{scherer13}. Figure~\ref{fig:theory}(a) shows the effective potential experienced by $\ket{\pm1}$ and the cylindrical box approximation. This motivates the simplified Hamiltonian for excitations of the system
  \begin{align}
    \hat{\Hex}&=\sum_{m=\pm1} \int d^2\mathbf{r}\, \dPsi_{m}^\dagger \left(-\frac{\hbar^2}{2m}\nabla^2+ U_1n_0 + q\right)\dPsi_{m}\nonumber\\
    &+ U_1 \int d^2\mathbf{r}\, n_0 \left( \dPsi_1^\dagger \dPsi_{-1}^\dagger + \dPsi_1 \dPsi_{-1} \right).
    \label{eq:hamiltonian2}
  \end{align}
The eigenstates of the cylindrical box potential are ~\cite{scherer10},
\begin{equation}
  \label{eq:besselmodes}
  \varphi_{nl} (\rho, \phi) = \frac{1}{\sqrt{\pi}\rTF J_{\abs{l}+1}(\beta_{nl})} J_{\abs{l}}\bigl(\beta_{nl} \frac{\rho}{\rTF}\bigl) e^{il\phi},
\end{equation}
where $\rTF$ is the Thomas-Fermi radius of the cloud, $J_{\abs{l}}$ are Bessel functions of the first kind, and $\beta_{nl}$ is the $n$'th zero of $J_{\abs{l}}$. The eigenstates $\varphi_{nl}$ have energies \mbox{$\epsilon_{nl}=\hbar^2\beta^2_{nl}/(2M\rTF^2)$.} 
The substitution of $\dPsi_{m}=\sum_{nl} \varphi_{nl}\hat{a}_{nlm}$ into Eq.~\eqref{eq:hamiltonian2} leads to the matrix elements 
$\braopket{\Bessel_{nl}}{n_0}{\Bessel_{n' l'}}$. Cylindrical symmetry of $n_0$ and orthogonality of $\varphi_{nl}$ for different $l$  
yields $\braopket{\Bessel_{nl}}{n_0}{\Bessel_{n' l'}}=\delta_{ll'}\braopket{\Bessel_{nl}}{n_0}{\Bessel_{n' l}}$. 
Additionally, a numerical evaluation of the matrix elements shows that $\braopket{\Bessel_{nl}}{n_0}{\Bessel_{n'l'}}\simeq\delta_{nn'}\delta_{ll'}\braopket{\Bessel_{nl}}{n_0}{\Bessel_{n l}}\equiv\delta_{nn'}\delta_{ll'}\expec{n_0}_{nl}$ is a good approximation. With this simplification, Eq.~\eqref{eq:hamiltonian2} may be put into a symmetric form and diagonalized through a Bogoliubov transformation. The excitation energies are 
\begin{equation}
  \label{eq:energies}
 \Einstab= \pm i\sqrt{\bigl( U_1 \expec{n_0}_{nl} \bigr)^2-\bigl(\epsilon_{nl} + U_1 \langle n_0 \rangle_{nl} + q\bigr)^2}.
\end{equation}
The eigenvalues $\Einstab$ are either real or imaginary depending on the interplay of $\epsilon_{nl}$, $U_1 \langle n_0 \rangle_{nl}$ and $q$: a real $\Einstab$ determines the standard phase evolution of an eigenstate; an imaginary value describes unstable evolution, i.e., growth or decay in the amplitude of an eigenstate. In the following, we focus on the unstable evolution and refer to  $E^\pm_{nl}/h$ as instability rates.

For our experimental parameters, the instability rates form a dense `forest' of overlapping resonances. 
The  density of \0 atoms sets the width and peak value of each resonance, and it is the extreme compression along the lattice symmetry axis that causes  many modes to overlap.  
For \mbox{$N_\mathrm{BEC}=4.2\times10^4$},
the central site has $9\times10^3$ atoms, and the peak value of the mean field repulsion  $U_1 n_0$ is  55~Hz (see Fig.~\ref{fig:theory}(a)).
 To illustrate how the modes change in shape as a function of $l$,  \figref{fig:theory}(b) shows  $E^\pm_{nl}$ for $n=1,\, l=0,1,\dots, 9$ in the central lattice site. The peak instability rate is reached by the mode $(1,0)$ due to its large overlap with  the BEC in \0. Figure~\ref{fig:theory}(c) shows the instability rates for all relevant modes in the range $q=0$ to $-500$~Hz: it is clear that a single $q$ value supports many unstable modes.
 
 We may quantify the multi-mode character of the spin dynamics by investigating the instability rates at a given $q$. We focus on the high $\abs{q}$ `free-space' regime. Figure~\ref{fig:theory}(d) shows the instability rates in a narrow interval around $q=-400$~Hz, for which 19 modes are unstable. 
We quantify multi-mode amplification  by the ratio $\delta E/E$,  
where $E/h$ is the mean value of the instability rate of the unstable modes, and $\delta E/h$ is the mean difference of instability rate between modes.
   For  $\delta E/E\ll1$, all modes grow in population almost equally on a timescale $(4\pi E/h)^{-1}$, as though degenerate in instability rate. 
In the case of the most unstable modes in \figref{fig:theory}(d), $\delta E/h\sim 2$~Hz and $E/h\sim30$~Hz, for which the  time scale for population growth is 2.7~ms, consistent with the experimentally measured value of 3.0~ms (see \figref{fig:spectrum}(b) and associated discussion). Only on a timescale $(4\pi\delta E/h)^{-1}\sim40$~ms will the evolution of different unstable modes become resolvable,  but this lies in the non-linear regime, well outside the 8~ms evolution time we employ (see Fig. 1(b)).

The expansion of excitations in $\ket{\pm1}$ in terms of states with angular momentum arises naturally from the cylindrical symmetry of the optical lattice, but this choice of basis is arbitrary.
In the `free-space' picture, two atoms from the stationary BEC undergo a momentum-conserving spin-changing collision: in the cylindrical basis, one spin state gains positive angular momentum $L$ while the other gains $-L$, and the total angular momentum remains zero.

We simulate the experimental results by forming a superposition state $\psi$ in each lattice site comprised of the set  $\{\nu\}$  of modes that have a finite instability rate at a given $q$. 
 In light of earlier work \cite{scherer10,scherer13}, we form a general superposition state  for one spin state (e.g., $\ket{+1}$) in each lattice site, consisting of positive and negative angular momentum components. 
The expansion coefficients  are sampled from appropriate probability distributions for each realization of the simulation~\cite{scherer13}.
The state is given by $\psi=\sum_{nl}^\nu c_{nl}\varphi_{nl}+ c_{nl}'\varphi_{n-l}$, where the complex amplitude is $c_{nl}=\sqrt{P_{nl}(t)}\rme^{i\theta}$, and $c_{nl}$ and $c'_{nl}$ are sampled separately. The first factor,  $\sqrt{P_{nl}(t)}$, is the square root of the probability 
to obtain $N$ atoms in mode $(n,l)$ in $\ket{\pm1}$; i.e., a two-mode  Fock state~\cite{scherer13}. 
The second factor, $\rme^{i\theta}$, establishes a random phase between $l$ and $-l$ modes, and between different lattice sites.
For each lattice site the BEC population is calculated and $\psi$ is  evolved for 20~ms in free space  \footnote{We model the transfer  to the optical lattice by assuming that the atoms are `frozen' by the lattice at $s=10$. 
The BEC   in the relaxed magnetic trap is taken to have a  Thomas-Fermi profile  along the lattice symmetry axis and we numerically integrate the profile over successive intervals of $\lambda/2$ to obtain the atom number in each site.}. The superposition across all lattice sites is then formed and the squared modulus is taken to obtain the density distribution. Finally, the simulated density distributions are convolved with the measured point spread function.

\begin{figure}[t]
  \includegraphics[width=8.6cm]{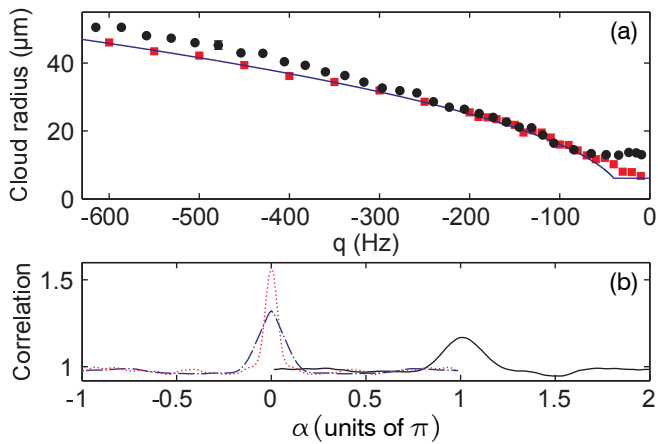}
\caption{(Color online) (a) Cloud radius after 20~ms time-of-flight.  (Black points)  experimental data, (red squares) simulations,  (blue line) ballistic expansion model  (offset to begin at the average value of the mean field repulsion across the populated lattice sites). (b) Density correlation function at $q=-297$~Hz (mean of 54 realizations for both simulations and experimental data): (black line) cross-correlation for  $\ket{\pm1}$ clouds, (blue dash-dot line) mean auto-correlation function for $\ket{\pm1}$ clouds,  (red dotted line) auto-correlation for  simulations.}
  \label{fig:ring-size-contrast}
\end{figure}

Examples of the simulated density distributions obtained following this procedure are shown in \figref{fig:images}(b). The simulated distributions  reproduce the transition from a density distribution with a central peak at low $\abs{q}$ to the striking ring structures observed in the experiment at high $\abs{q}$.
This transition can be understood from the mode structure in Figs.~\ref{fig:theory}(b) and (c): the dominant $(1,0)$ mode extends from $q\approx0$ to $q\approx -90$~Hz, meaning that $\psi$ in this interval will have non-zero density in the center of the cloud, even after time-of-flight. 
For larger $\abs{q}$, modes without population in the center and a range of angular momenta predominate in the experimental and simulated density distributions. While the simulations capture the overall ring structure well, 
it is clear that they  show more azimuthal structure than  the observed density distributions.

To make a  quantitative comparison of the simulated and experimental density distributions, we first investigate the size of the clouds. Figure \ref{fig:ring-size-contrast}(a) shows the cloud radius $\expec{\rho}$ of the experimental and simulated density distributions.
  The box model simulations show good  agreement with the experimental cloud size, with deviations arising for low and high $\abs{q}$ where the cylindrical box approximation breaks down. At low $\abs{q}$, this arises from the $\sim50$~Hz repulsive bump $U_1 n_0$ (see Fig. 2(a)), and at high $\abs{q}$, the box potential underestimates the energy imparted to the scattered atoms due the finite slope of the walls. One can obtain a simple estimate of the cloud radius in the high $\abs{q}$ regime  using the `free space' picture:  if each wave packet gains $p_\textrm{rms}=\sqrt{2Mq}$, the position of an atom assuming  ballistic expansion is given by $\expec{\rho} = \sqrt{\rho_0^2 + (p_\textrm{rms}t/M)^2}$, where ${\rho_0}$ is the in-trap radius and $t$ is the time-of-flight. 

To study the structure around the circumference of the experimental and simulated density distributions,  we employ an angular density correlation function.
Momentum conservation in the collision process requires that a density peak at  angle $\theta$ in a $\ket{\pm1}$ cloud leads to an anti-correlated density peak located at \mbox{$\theta^\prime = \theta - \pi$} in the $\ket{\mp1}$ component (see 
\figref{fig:images}(a) for $q=-382$~Hz). 
We define the  correlation function \mbox{$c(\alpha) = \langle \tilde{n}_{-1}(\theta) \tilde{n}_{+1}(\theta - \alpha) \rangle / [\langle \tilde{n}_{-1}\rangle \langle \tilde{n}_{+1}\rangle]$}, where $\tilde{n}_{\pm1}$ are the two angular density distributions
and the angled brackets denote the mean over $\theta$. The  correlation function for $q=-297$~Hz is shown in Fig.~\ref{fig:ring-size-contrast}(b) and exhibits the expected anti-correlation between the two clouds.  In order to analyze the $l$-mode composition of the states generated by the spin dynamics, we also show the  auto-correlation function for the experimental and simulated density distributions. A comparison of these indicates that only a subset of the energetically allowed $l$ modes contribute to the observed experimental states.
Nonetheless, the auto-correlation function  confirms that several $l$ states contribute to the observed distributions because $c(\alpha)$ is approximately flat  away from the peak at $\alpha=0$ due to the destructive interference of several frequency components; in the `free space' picture, this amounts to  random  spontaneous symmetry breaking of the polar symmetry.

In conclusion, we have investigated spin dynamics in a 2D quantum gas and have observed qualitatively new features such as  ring-shaped density distributions  in time-of-flight. A theoretical analysis of the  spin  excitation modes in the system provided a qualitative understanding of the ring-shaped clouds and showed  their angular structure is due to the interference of several modes. 
In future  work, we will gain a full understanding of the unstable spin  modes  by studying the dynamics in a single site of the optical lattice. Finally, tailoring the trapping potential further will enable the preparation  of anti-correlated wave-packets of quantum degenerate atoms with the aim of producing an atom source with non-local EPR entanglement.

\begin{acknowledgments}
We acknowledge support from the Danish National Research Foundation, the Danish Council for Independent Research, the Lundbeck Foundation, the DFG (SA1031/6), the Cluster of Excellence QUEST and the German-Israeli Foundation.
\end{acknowledgments}
%\bibliography{../spinor}
\bibliography{superpositions_spinor_CutResub}
\end{document}